\documentclass[onecolumn,prx,longbibliography,nofootinbib]{revtex4-2}

\usepackage[dvips]{graphicx} % for figures
\usepackage{amsfonts,amscd,amsmath,amsthm}
\usepackage{enumerate}
\usepackage{epsfig}
\usepackage{subfigure}
\usepackage{xcolor}
\usepackage[colorlinks = true]{hyperref}
\usepackage{physics}
\usepackage{epstopdf}
\usepackage{framed}
\usepackage{multirow}
\usepackage{color}
\usepackage{longtable}
\usepackage{comment}
\usepackage[ruled,vlined]{algorithm2e}
\usepackage[most]{tcolorbox}

\graphicspath{{./figure/}}

\usepackage{tikz}
\usetikzlibrary{tikzmark, calc, fit, positioning}
\usetikzlibrary{shapes}
\usetikzlibrary{quantikz}

\newtheorem{observation}{Observation}

\newtcolorbox[auto counter]{mybox}[2][]{
	enhanced,
	breakable,
	colback=blue!5!white,
	colframe=blue!75!black,
	fonttitle=\bfseries,
	title=Box \thetcbcounter: #2,#1
}

\begin{document}

%\title{Beam-splitting Attacks on Phase-Matching Quantum Key Distribution}
\title{Source-Replacement Model for Phase-Matching Quantum Key Distribution}
%% For REVTeX it is possible to automate superscript and e-mail callouts with the superscriptaddress option; see REVTeX4 documentation.

\author{Yizhi Huang}
\affiliation{Center for Quantum Information, Institute for Interdisciplinary Information Sciences, Tsinghua University, Beijing 100084, China}
\author{Zhenyu Du}
\affiliation{Center for Quantum Information, Institute for Interdisciplinary Information Sciences, Tsinghua University, Beijing 100084, China}
\author{Xiongfeng Ma}
\email{xma@tsinghua.edu.cn}
\affiliation{Center for Quantum Information, Institute for Interdisciplinary Information Sciences, Tsinghua University, Beijing 100084, China}

\begin{abstract}
%Quantum key distribution has emerged as a promising solution for constructing secure communication networks, offering information-theoretic security guaranteed by the principles of quantum mechanics. One of the most advanced quantum key distribution protocols to date is the phase-matching protocol, whose security was initially proven using a method called symmetry-protected privacy. In this work, we reexamine the security of this protocol, returning to a derivation based on the entanglement picture and arriving at conclusions consistent with the original proof. This provides a new perspective on the security of the PM scheme. As an application of this perspective, we a novel beam-splitting attack scheme. By employing the entanglement-based analysis, we obtain an lower bound on the phase error rate under this attack, further demonstrating the tightness of the security analysis method in terms of key rate.

Quantum key distribution has emerged as a promising solution for constructing secure communication networks, offering information-theoretic security guaranteed by the principles of quantum mechanics. One of the most advanced quantum key distribution protocols to date is the phase-matching protocol. Its security was initially established using an abstract method known as symmetry-protected privacy. In this study, we reevaluate the security of the phase-matching protocol using an intuitive source-replacement model, and we arrive at conclusions that align with the original proof. This model provides a fresh perspective on the protocol's security. As an application of this approach, we introduce a beam-splitting attack scheme. Leveraging the source-replacement model, we derive a lower bound on the phase error rate under this attack, further underscoring the robustness of our security analysis method.\\
%, whose security was initially proven using an abstract method called symmetry-protected privacy. In this study, we revisit the security of the phase-matching protocol, utilizing an intuitive source-replacement model and arriving at conclusions consistent with the original proof. This model provides a new perspective on the security of the phase-matching protocol. As an application of this perspective, we proposed a beam-splitting attack scheme. By employing the source-replacement model, we obtain a lower bound on the phase error rate under this attack, further demonstrating the tightness of the security analysis method.\\
\end{abstract}

%\begin{keywords}
%quantum key distribusion, phase-matching scheme, pseudo-Fock state, beam-splitting attack
%\end{keywords}

\maketitle %% required
%\tableofcontents
%\clearpage

\section{Introduction}\label{sc:intro}
Quantum key distribution (QKD)~\cite{bennett1984quantum,ekert1991Quantum} is currently one of the most successful applications in quantum information science. It allows remote communication parties, Alice and Bob, to establish a secure key by leveraging the principles of quantum mechanics. The information-theoretical security of QKD has been proven theoretically since the end of last century based on the quantum bit and phase error correction~\cite{lo1999Unconditional,shor2000Simple,gottesman2004security,koashi2009simple} or the entropic approach~\cite{devetak2005distillation,renner2008security}. Here, bit error rates can tell differences between the key strings held by Alice and Bob, while phase error rates can provide an upper bound on the potential information leakage. Therefore, by performing bit and phase error correction, Alice and Bob can share two consistent and secure key strings.
The information-theoretic security of QKD enables it to withstand attacks from quantum computers, which has attracted significant interest from researchers. 
In 2012, Lo \emph{et al.}~proposed measurement-device-independent (MDI) QKD \cite{lo2012Measurement} further propelling advancements in this fields. The security of MDI QKD requires no assumption on how the measurement site performs measurement and announcement, making it naturally immune to all detection attacks. Meanwhile, it helps reduce the number of trusted nodes and makes quantum communication networks more implementable.
%Later, another type of MDI QKD, the twin-field scheme \cite{lucamarini2018overcoming}, is proposed to achieve quadratic improvement in terms of key rate \cite{Ma2018phase,tamaki2018information,wang2018twin,lin2018simple}, surpassing the linear bound \cite{takeoka2014fundamental,pirandola2017fundamental}. 
Later, another type of MDI QKD, known as the twin-field scheme \cite{lucamarini2018overcoming}, was introduced with the aim of achieving a quadratic improvement in key rate \cite{Ma2018phase, tamaki2018information, wang2018twin, lin2018simple}.	This approach successfully allowed for a breakthrough in the key rate beyond the linear bound \cite{takeoka2014fundamental, pirandola2017fundamental} without the need for quantum relays.
Recently, a protocol called mode-pairing QKD has adopted innovative pairing ideas, achieving a quadratic improvement in the key rate while maintaining a relatively simple implementation difficulty as conventional MDI setups \cite{zeng2022mode,zhu2023experimental,zhou2023experimental}. With the introduction of various protocols, the practicality, communication distance, and speed of QKD have made significant advancements in recent years. For reviews of this subject, one may refer to \cite{Xu2020Secure,portmann2022security}.

Among the various QKD protocols developed, the phase-matching (PM)scheme \cite{Ma2018phase} has garnered significant attention due to its robustness and efficiency. The security of the PM scheme was originally established using a method known as symmetry-protected privacy \cite{lin2018simple,Zeng2019Symmetryprotected}. Unlike traditional complementary-based security proofs, this method utilizes the symmetry of encoding to establish security, leveraging the parity properties of the corresponding state space to derive the phase error rate. The advantage of this proof technique lies in its analysis being entirely based on encoding operations, independent of the exact form of the source and measurements. It provides a straightforward and convenient framework for analyzing the security of QKD protocols, particularly MDI QKD protocols. 
However, this simple and highly abstract nature also poses certain problems. The symmetry-protected privacy method becomes less applicable in providing specific security discussions for different protocols. Furthermore, it often limits the analysis to the encoding operations carried out by the communicating parties, making it difficult to draw conclusive results about the protocol's performance in the presence of potential eavesdroppers when an attack occurs.
 
In this work, we attempt to reexamine the security of the PM scheme. We start by considering the source-replacement approach, which is more concrete and easier to comprehend. This approach was first introduced in \cite{shor2000Simple}, and its name comes from \cite{Ferenczi2012symmetries}. Using an entanglement-based protocol equivalent to the original protocol, we introduce virtual CNOT gates, quantum Fourier transforms, and photon number measurements to form a pseudo-Fock state and show how to simultaneously get the total photon number and the random phase difference. Finally, based on the original definition, we establish the relationship between photon numbers and phase errors. 
By approaching the problem from this different perspective, we arrive at the same conclusion as the symmetry-protected privacy method: quantum states with an odd total photon number result in phase errors, while states with an even total photon number do not. 

%xxxxx source-replacement (ref in mode-pairing paper) first introduced in Shor-Preskill, name comes from Lutkenhaus' paper

By delving into the source-replacement analysis, we aim to provide a new viewpoint that complements the original proof of security. Our analysis reaffirms the protocol's security and provides valuable insights into its underlying mechanisms. In addition, we introduce a beam-splitting attack scheme that poses a potential threat to the PM scheme. By using the source-replacement approach, we derive an upper bound on the phase error rate, quantifying the attack's impact on the protocol's performance. By analyzing the simulation results, we demonstrate that the phase error rate provided by the security proof is very close to the one introduced by the beam-splitting attack. This finding indicates that the lower bound on the key rate provided by security analysis is already highly tight, leaving little room for further improvement. Our proposed analysis establishes a direct connection between the attack and the quantum phase error rate, elucidating the security of the PM scheme. 

The rest of the content of this paper is as follows. In Sec.~\ref{sc:preliminary}, we review the process of the phase-matching scheme and introduce the sketch of symmetry-protected security proofs. In Sec.~\ref{sc:SourceRep}, we introduce our security proof based on the source-replacement approach and derive the phase error rate. In Sec.~\ref{sc:attack}, we present the attack strategy, analyze the lower bound of the phase error rate resulting under this attack using the source-replacement model, and simulate the disparity between this lower bound and the phase error rate provided by security analysis. Finally, in Sec.~\ref{sc:conclusion}, we provide a conclusion and outlook of our work.

\section{Preliminary}\label{sc:preliminary}
In this section, we will introduce the specific steps of the PM scheme and its characteristics. We will also provide a brief overview of the core principles and main conclusions of the symmetry-protected privacy security proof method.
\subsection{Phase-matching scheme}\label{sc:PM}
Firstly, we introduce the phase-matching QKD scheme. The core idea of this scheme is that Alice and Bob encode their respective key information into the phase of individual optical pulses, with key bits 0 and 1 corresponding to phase values of $0$ and $\pi$, for instance. Subsequently, they send their pulses to Charlie for single-photon interference. By analyzing the interference outcomes, they can ascertain the degree of phase matching between their encoded phases. This process, conducted through a single optical mode, establishes the connection between the key information of both parties. A more detailed description of the protocol steps is shown in Box \ref{box:PMQKD}.
\begin{mybox}[label={box:PMQKD}]{Phase-matching QKD scheme \cite{Ma2018phase}}%\label{box:BB84}
	\begin{enumerate}[(1)]
		\item 
		State Preparation: In the $i$-th round, Alice prepares coherent state $\ket{\alpha_i}=\ket{\sqrt{\mu_a^i}e^{ \mathrm{i} (\pi\kappa^i_a+\phi^i_a)}}$ on optical mode $A_i$, where $\mu_a^i$ is the intensity, $\kappa^i_a$ is Alice's raw key bit which chosen from $\{0,1\}$ randomly, and phase $\phi^i_a$ is uniformly chosen from $[0,2\pi)$. Similarly, Bob randomly chooses $\kappa^i_b$ and $\phi_b^i$, and prepares $\ket{\beta_i}=\ket{\sqrt{\mu_b^i}e^{ \mathrm{i} (\pi\kappa^i_b+\phi^i_b)}}$ on mode $B_i$.
		\item
		Measurement: Alice and Bob send their optical modes $A_i$ and $B_i$ to an untrusted party, Charlie, who is supposed to perform single-photon interference measurement and records the clicks of detectors $L$ and/or $R$.
		\item %\label{StepAnnounce}
		Announcement: For all rounds with successful detection, Charlie announces the $L$ and $R$ detection results.
		Then, Alice and Bob announce the random phases, $\phi^i_a$ and $\phi_b^i$, of these rounds.
		\item %\label{StepParaEst}
		Key mapping: Alice and Bob repeat the above steps for $N$ rounds. For those rounds with successful detections, Alice and Bob compare their encoded random phases $\phi^i_a$ and $\phi_b^i$. If $\abs{\phi^i_a-\phi^i_b}=0$ or $\pi$, Alice and Bob keep the corresponding $\kappa_a^i$ and $\kappa_b^i$ as their raw key bits in this round, respectively. Otherwise, the discard their raw key bits. Additionally, Bob flips his key bit $\kappa_b^i$ if Charlie’s announcement was an $R$ click and he also flip his key bit if $\abs{\phi^i_a-\phi^i_b}=\pi$.
		\item %\label{StepKeymeasure}
		Parameter estimation: For all the left raw key bits, Alice and Bob analyze the gains $Q_\mu$ and quantum bit error rates $E_Z$. Here, the gain $Q_\mu$ is defined as the successful detection probabilities of the signal states with intensity $\mu$, and the quantum bit error rates $E_Z$ can be obtained through random sampling. Then they estimate the phase error rate $e_p$ use the method in \cite{Zeng2019Symmetryprotected}.
		\item %\label{StepClassical}
		Post-processing: Alice and Bob reconcile the key via an classical channel. They then perform privacy amplification according to $e_p$ to get the final key bits.
	\end{enumerate}
\end{mybox}

The key observation of the phase-matching scheme is that by utilizing the result of single-photon interference, Alice and Bob can distill raw key bits from a single detection. The probability of a successful detection event is proportional to $\sqrt{\eta}$ when either Alice's or Bob's photon causes a detection click, where $\sqrt{\eta}$ is the transmittance from Alice or Bob to Charlie and here we suppose the channel is symmetric. Therefore, the key rate performance of the phase-matching scheme is $O(\sqrt{\eta})$. This is a quadratic improvement compared to conventional MDI schemes whose key rate performance is $O(\eta)$ due to the requirement of coincidence detection.

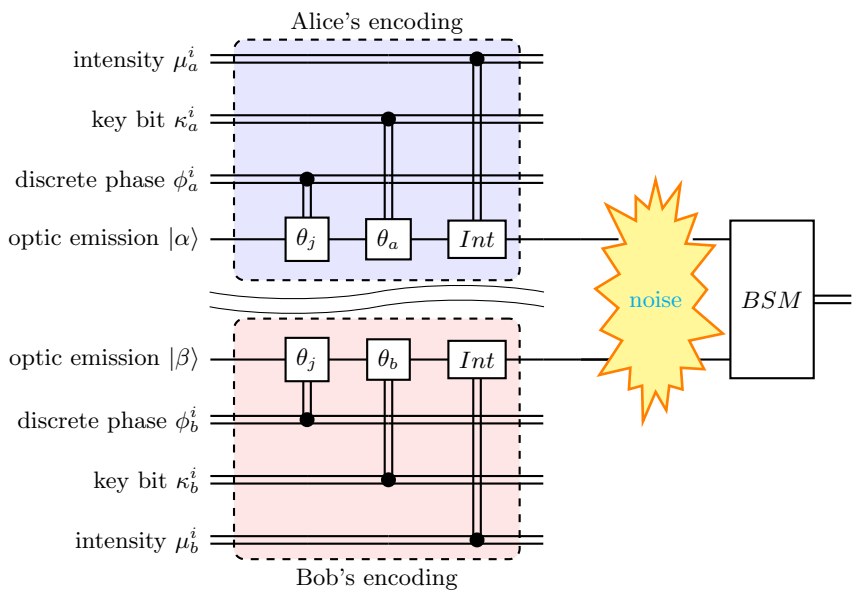
\begin{figure}[hbtp!]
	%\begin{tikzpicture}
	%\node[scale=.6] {
		\begin{quantikz}[row sep={.8cm,between origins}]
			\lstick{intensity $\mu_a^i$}& \cw \gategroup[4,steps=4,style={dashed, rounded corners,fill=blue!10, inner xsep=2pt}, background]{{Alice's encoding}}  & \cw & \cw & \cwbend{3} &\cw  \\
			\lstick{key bit $\kappa_a^i$} &\cw  & \cw & \cwbend{2} & \cw  & \cw\\
			\lstick{discrete phase $\phi_a^i$} & \cw & \cwbend{1} & \cw  & \cw & \cw \\
			\lstick{optic emission $\ket{\alpha}$}  & \qw & \gate{\theta_j} & \gate{\theta_a} & \gate{Int}  & \qw & \qw & \gate[3,nwires={2},style={starburst,fill=yellow!50,draw=orange,line width=1pt,inner xsep=-4pt,inner ysep=-5pt}, label style=cyan]{\text{noise}} & \gate[3,nwires={2}]{BSM} \\
			\wave &&&&&&&&& \cw \\
			\lstick{optic emission $\ket{\beta}$} &\qw \gategroup[4,steps=4,style={dashed, rounded corners,fill=red!10, inner xsep=2pt}, background,label style={label position=below,anchor=north,yshift=-0.2cm}]{{Bob's encoding}} & \gate{\theta_j} & \gate{\theta_b} & \gate{Int} & \qw & \qw & \qw &   \\
			\lstick{discrete phase $\phi_b^i$}&\cw & \cwbend{-1} & \cw & \cw & \cw  \\
			\lstick{key bit $\kappa_b^i$} & \cw & \cw & \cwbend{-2} & \cw& \cw \\
			\lstick{intensity $\mu_b^i$} & \cw & \cw & \cw & \cwbend{-3} & \cw
		\end{quantikz}
		%};
	%\end{tikzpicture}
	\caption{Encoding circuit of PMQKD. This circuit is symmetric for Alice and Bob, and here, we only introduce the systems and operations on Alice's side. The systems and operations on Bob's side are analogous to those of Alice. Systems $A$ and $B$ are optical modes initialized to coherent states $\ket{\alpha}$ and $\ket{\beta}$, respectively. Alice uses quantum random number generator to generate $\mu_a^i,\kappa_a^i,\phi_a^i$ according to the protocol, and then she uses the classical control-phase operation to encode the discrete phase and key information according to $\phi_a^i$ and $\kappa_a^i$. She also uses the random number $\mu_a^i$ to control the intensity of $\ket{\alpha}$ to carry out the decoy state estimation. 
	%Here, we assume Alice and Bob perform the vacuum+weak decoy state protocol \cite{Ma2005practical} for intensity control.
 } \label{fig:PMencodingClassical}
\end{figure}

To provide a clearer illustration of the encoding, we show the encoding process using quantum circuit notation in Figure \ref{fig:PMencodingClassical}. In this representation, we introduce three classical random bits at each side for classical control to encode the key bits, random phases, and intensities on the optical mode. The control parameters, $\mu_a^i,\mu_b^i,\kappa_a^i,\kappa_b^i,\phi_a^i,\phi_b^i$, are random numbers coming from quantum random number generators according to the scheme.

\subsection{Symmetry-protected security}\label{sc:symmetrypro}
Unlike the BB84 scheme or the conventional MDI scheme, the PM scheme utilizes only one basis for key generation and parameter estimation. Consequently, applying the conventional security proof based on basis complementarity \cite{koashi2009simple} directly to the phase-matching scheme becomes challenging. To address this issue, an approach known as symmetry-protected privacy was introduced in \cite{Zeng2019Symmetryprotected}. This method establishes security proof based on a different principle, namely encoding symmetry. The key encoding operation on the initial state can always be described in the following form:
\begin{equation}
	U_{AB}(\kappa_a,\kappa_b) = U_A(\kappa_a)\otimes U_B(\kappa_b).
\end{equation}
By leveraging the its unitarity, the Hilbert space can be divided into two distinct subspaces. The first is the even space $\mathcal{H}^{\text{even}}$, which consists of even states that are eigenstates of $U$ with an eigenvalue of $+1$. The second is the odd space $\mathcal{H}^{\text{odd}}$, which comprises odd states that are eigenstates of $U$ with an eigenvalue of $-1$. For an initial state $\rho_{AB}$, we can define the encoded state $\rho'_{AB}$ as 
\begin{equation}
	\rho_{AB}' (\kappa_a,\kappa_b)= \left[U_A(\kappa_a)\otimes U_B(\kappa_b)\right]\rho_{AB}\left[U_A(\kappa_a)\otimes U_B(\kappa_b)\right]^\dagger.
\end{equation}

Intuitively, if the initial state $\rho_{AB}$ is a mixture of states solely from one of the two subspaces, then the four possible encodings will produce the same encode state in pairs, i.e., $\rho_{AB}' (0,0)=\rho_{AB}' (1,1)$ and $\rho_{AB}' (1,0)=\rho_{AB}' (0,1)$.
Consequently, Eve cannot determine the specific encoding operations employed by Alice and Bob when she obtains the encoded state. Therefore, Eve at most can know whether Alice and Bob's encoding is the same, and she can not get the specific encoding information. In this scenario, the phase error rate is zero.
However, unfortunately, both Alice and Bob are also unaware of each other's encoding operations. As a result, the bit error rate will be 50\%, and the final key rate will be 0. In practice, the initial state $\rho_{AB}$ is a mixture of both even and odd states, 
\begin{equation}
	\rho_{AB} = p_{\text{even}}\rho_{\text{even}} + p_{\text{odd}}\rho_{\text{odd}},
\end{equation}
where $\rho_{\text{even}}=\sum_{\rho_i\in \mathcal{H}_{even}} p_i \rho_i$, $\rho_{\text{odd}}=\sum_{\rho_j\in \mathcal{H}_{odd}} p_j \rho_j$. When adopting such a mixture state, Eve can, to some extent, distinguish different encoding outcomes by observing the relative phase changes on the purified state $\ket{\Psi}_{ABE}$ of $\rho_{AB}$, where system $E$ is under Eve's control, leading to a non-zero phase error rate.

In the symmetry-protected security proof, it is further demonstrated that in this particular scenario, the phase error rate, denoted as $e_p$, corresponds to the proportion of odd or even components that contribute to effective detection. These components are represented by $q_{\text{odd}}$ and $q_{\text{even}}$ respectively, and it is important to note that they differ from the previously mentioned $p_{\text{odd}}$ and $p_{\text{even}}$. The variable $p$ denotes the components in the initial state, while the variable $q$ pertains to the post-selected states that have been successfully detected. The relationship between these two kinds of quantities can be expressed as 
\begin{equation}
\begin{split}
q_{\text{odd}} &= p_{\text{odd}}\dfrac{Y_{\text{odd}}}{Q}, \\
q_{\text{even}} &= p_{\text{even}}\dfrac{Y_{\text{even}}}{Q},
\end{split}
\end{equation}
where $Y_{\text{odd/even}}$ and $Q$ represent the successful detection probabilities of $\rho_{{\text{odd/even}}}$ and $\rho_{AB}'$ respectively. 

In the phase-matching scheme, the odd and even states correspond to Fock states with an odd or even number of photons, respectively. Therefore, the phase error in the phase-matching scheme is 
\begin{equation}\label{eq:epupper}
	e_p \leq q_{\text{even}} = 1 - \sum_{k} q_{2k+1} \equiv e_p^u,
\end{equation}
where $q_k$ is the fraction of detection when Alice and Bob send out $k$-photon signals in all. For ideal case where $\mu_a=\mu_b=\mu$, we can employ the following formulas to calculate $q_{even}$ by 
\begin{equation}
	q_k = \frac{Y_k (2\mu)^k e^{-2\mu}}{k! Q_{\mu}},
\end{equation}
where $Y_k = 1-(1-\eta)^k$ and $Q_{\mu} = 1-e^{-2\eta\mu}$ are successful detection probabilities of $k$-photon state and coherent state with intensity $\mu$, respectively. The key rate of the phase-matching scheme is \cite{Ma2018phase}
\begin{equation}
R \geq Q_\mu\left[1-h\left(e_p\right)-fh(E_{\mu})\right],
\end{equation}
where $E_{\mu}$ is the bit error rate that can be directly obtained from experimental data, $h(x)=-x\log(x)-(1-x)\log(1-x)$ is the binary entropy function, and $f$ is the error correction efficiency. Note that in ref.~\cite{Zeng2019Symmetryprotected}, the security of QKD under symmetric encoding is also analyzed using the standard phase-error-correction method, proving that the upper bound on the phase error rate provided by the symmetry-protected method is secure.

\section{Source-replacement security proof}\label{sc:SourceRep}
The symmetry-protected privacy method has been rigorously established for security proof, characterized by a highly abstract and overarching analytical process. This approach allows the analytical framework to extend beyond the PM scheme, encompassing other protocols employing symmetric encodings. However, this abstraction complicates the intuitive understanding of the security of specific protocols. For instance, unlike conventional practices, where the phase error rate is obtained through measurements on qubits held by Alice and Bob, this method employs a different analytical process. Therefore, while this method provides an upper bound on the phase error rate based on photon components for the PM scheme, it cannot explain the underlying physical connection between these two factors in a concrete way. 
Furthermore, when applying the symmetry-protected privacy analysis, one do not need to consider the potential attack an eavesdropper may carry out. Although this simplifies the analysis, it also limits the method's efficacy when evaluating the impact of specific attacks on the protocol. As a result, when attempting to dissect the influence of a particular attack on the protocol, this analytical approach may fall short.

%xxxxx usually lots of random encoding info, such as basis, phase, to protect key bits. can be regarded as entanglement, replaced by classical operations. in principle, entanglement distillation, if forbidden, fail

In order to offer a more concrete and physically intuitive security proof tailored to the PM scheme, we have undertaken a reexamination of the protocol's security proof. Employing a source-replacement approach, we revert to the initial entanglement-based framework to define the phase error rate and derive conclusions consistent with those of the symmetry-protected privacy method. 
%The entanglement-based encoding circuit of the PM scheme is shown in Fig.~\ref{fig:PMencoding}. 

%After Alice and Bob send their optical modes to Charlie, Charlie will perform the bell-state measurement and announce the result. Then, Alice and Bob perform $Z$-basis measurement on systems $A_i$ and $B_i$ to get the encoded key bits, discrete phase, and intensity for postprocessing.

%In practice, Alice and Bob estimate $E_Z$ by random sampling on their key bits and comparing their keys. However, they cannot measure $X$ basis directly to estimate $e_p$. Instead, they perform measurements on $A_0, A_2$ and $B_0, B_2$ to estimate $e_p$ with the decoy state method, as shown in Figure \ref{fig:PMencoding} and the following sections. Here, we assume Alice and Bob perform the vacuum+weak decoy state protocol \cite{Ma2005practical} for intensity control. 

%We prove the security in the entanglement picture. That is, all operations are performed in a coherent way. We show that phase error is 1 when the total photon number in $A$ and $B$ is odd, and phase error is 0 when the total photon number is even. 

\subsection{Virtual operation and measurement}
Firstly, we attempt to transform the circuit in Figure \ref{fig:PMencodingClassical} into an entanglement-based protocol circuit by using the source-replacement approach. In general, the control parameters can be seen as $Z$-basis measurement results of $X$-basis eigenstates of Hilbert spaces with different dimensions. Taking Alice's side as an example, we introduce ancillary systems $A_0$ and $A_1$ to encode the discrete phase and the key information. They will be initialized to the eigenstates $\ket{+_d}$ of the Pauli $X$ operator in $d$-dimension according to the encoding requirements and $\ket{+}$, respectively. For example, if Alice and Bob choose to encode 16 random phases, i.e., $\phi_a^i\in \{\dfrac{2j\pi}{16}\}, j = 0,1,2,\cdots,15$ in Box \ref{box:PMQKD}, then the system $A_0$ will be initialized to $\ket{+_{16}}$. The $Z$-basis measurement results on these two systems act as $\phi_a^i$ and $\kappa_a^i$ and control the random phase encoding and the key bit encoding, respectively.
The situation with the ancillary system $A_2$ which controls the intensity is relatively complex. Here, we'll provide a simple example. We suppose the protocol involves only three types of light intensities: vacuum state intensity, decoy state intensity, and signal state intensity, with equal probabilities of being sent. In this case, system $A_3$ would be initialized to a 3-dimensional $X$-basis eigenstate $\ket{+_3}$. After performing a $Z$-basis measurement on it, the three possible outcomes will control the intensities of the pulses.

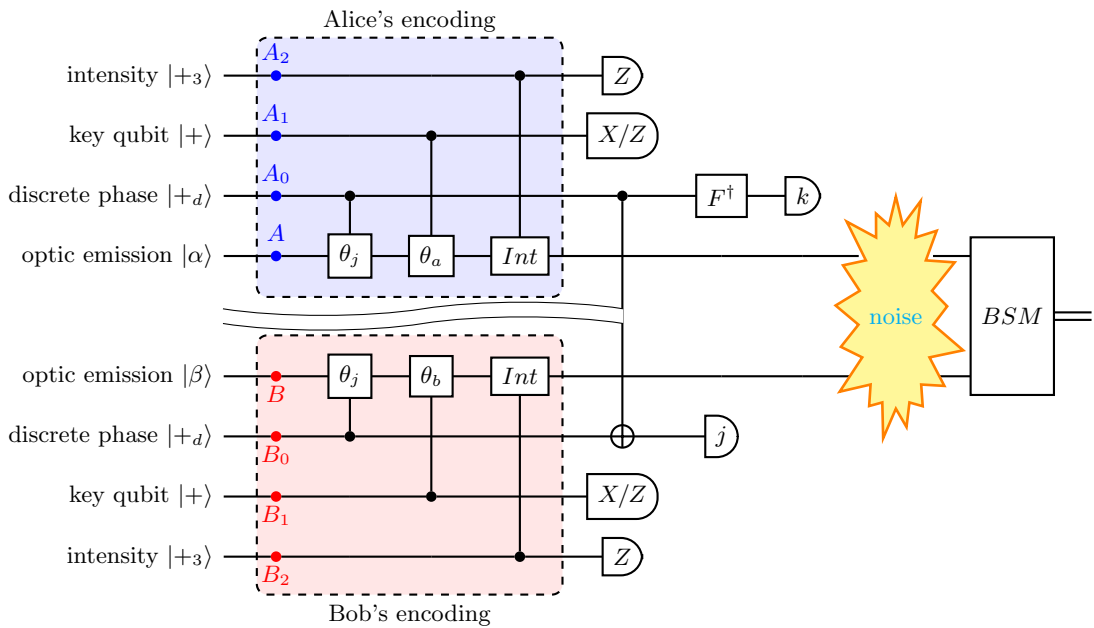
\begin{figure}[hbtp!]
	%\begin{tikzpicture}
	%\node[scale=.6] {
		\begin{quantikz}[row sep={.8cm,between origins}]
			\lstick{intensity $\ket{+_3}$} & \phase[blue,label position=above]{A_2} \gategroup[4,steps=4,style={dashed, rounded corners,fill=blue!10, inner xsep=2pt}, background]{{Alice's encoding}} & \qw & \qw & \ctrl{3} & \meterD{Z}  \\
			\lstick{key qubit $\ket{+}$} & \phase[blue,label position=above]{A_1} & \qw & \ctrl{2} & \qw & \meterD{X/Z}  \\
			\lstick{discrete phase $\ket{+_d}$} & \phase[blue,label position=above]{A_0} & \ctrl{1} & \qw & \qw & \ctrl{4} & \gate{F^\dagger} & \meterD{k} \\
			\lstick{optic emission $\ket{\alpha}$} & \phase[blue,label position=above]{A} & \gate{\theta_j} & \gate{\theta_a} & \gate{Int} & \qw & \qw & \qw & \gate[3,nwires={2},style={starburst,fill=yellow!50,draw=orange,line width=1pt,inner xsep=-4pt,inner ysep=-5pt}, label style=cyan]{\text{noise}} & \gate[3,nwires={2}]{BSM} \\
			\wave &&&&&&&&&& \cw \\
			\lstick{optic emission $\ket{\beta}$} & \phase[red,label position=below]{B} \gategroup[4,steps=4,style={dashed, rounded corners,fill=red!10, inner xsep=2pt}, background,label style={label position=below,anchor=north,yshift=-0.2cm}]{{Bob's encoding}} & \gate{\theta_j} & \gate{\theta_b} & \gate{Int} & \qw & \qw & \qw &  & \\
			\lstick{discrete phase $\ket{+_d}$} & \phase[red,label position=below]{B_0} & \ctrl{-1} & \qw & \qw & \targ{} & \meterD{j} \\
			\lstick{key qubit $\ket{+}$} & \phase[red,label position=below]{B_1} & \qw & \ctrl{-2} & \qw & \meterD{X/Z} \\
			\lstick{intensity $\ket{+_3}$} & \phase[red,label position=below]{B_2} & \qw & \qw & \ctrl{-3} & \meterD{Z}
		\end{quantikz}
		%};
	%\end{tikzpicture}
	\caption{Encoding circuit of source-replaced PM scheme. Similar to the circuit in Fig.~\ref{fig:PMencodingClassical}, the ancillary systems $A_0, A_1, A_2, B_0, B_1, B_2$ are used to encode the random phases, key bits, and intensities. In this picture, the ancillary systems are measured after the encoding operations, allowing Alice and Bob to employ additional operations beyond simple $Z$-basis measurements to estimate parameters like photon number and phase differences. The quantum inverse Fourier transform $F^\dagger$, high-dimensional CNOT operations, and $X$-basis measurements depicted in the figure are virtual operations not part of the original protocol.} \label{fig:PMsecurity}
\end{figure}

According to \cite{Liu2022classically}, the quantum phase control gates act as classically replaceable operations gates, enabling measurements to be performed prematurely and replaced with classical control. Utilizing this property, we can represent the measurement operations as the final steps performed by Alice and Bob, as depicted in Figure \ref{fig:PMsecurity}. In this picture, Alice and Bob initially initialize the ancillary systems into $\ket{+}$ states in different-dimensional Hilbert spaces, as required by the protocol. These ancillary systems serve as control bits, and they utilize quantum control operations to encode the optical modes before transmitting them to the measurement end. Finally, Alice and Bob perform $Z$-basis measurements on these ancillary systems to obtain the encoded key bits, random phases, and intensities in this round. 

The control operations entangle the ancillary systems of Alice and Bob with the optical modes they send to Charlie. Subsequently, their optical modes undergo measurements performed by Charlie, resembling an entanglement swapping process. This results in Alice and Bob sharing imperfect entanglements. Consequently, by replacing the source with a quantum encoded one, the entire protocol is transformed into an entanglement-based protocol. 
With the entanglement-based picture, we now try to derive the phase error rate of the PM scheme. In the following discussion, we will not focus on system $A_3$ and intensities of the optical mode, as varying light intensities only result in different photon number distributions and do not affect our subsequent derivations. According to the original security proof based on the quantum phase error \cite{lo1999Unconditional}, we need to estimate the quantum bit error rate and the quantum phase error rate for error correction in order to distill perfect entangled pairs between Alice and Bob and ensure the security of the key bits. Typically, the quantum bit error rate can be directly calculated from results obtained in practical experiments. Therefore, we will primarily focus on estimating the quantum phase error rate.

As we employ $Z$-basis measurements on systems $A_1$ and $B_1$ for key bits in PM scheme, the quantum phase error rate is naturally defined as the probability that Alice and Bob obtain inconsistent results when performing $X$-basis measurements on these two systems.

Furthermore, to obtain conclusions consistent with the symmetry-protected privacy method, specifically that the quantum phase error rate of the protocol is bounded by the total photon number distribution of the optical pulses sent by Alice and Bob in a single round, we introduce an additional virtual CNOT operation between systems $A_0$ and $B_0$. We replace the measurements of random phases on system $A_0$, acting as control bits, with a virtual quantum Fourier transform followed by a photon number measurement. In the end, the corresponding quantum circuit diagram for the protocol is depicted as Figure \ref{fig:PMsecurity}.

%According to \cite{Liu2022classically}, all the $Z$-basis measurement at the users' side can be exchanged with the control operations. Then the measurement results will become random numbers, and the quantum control operations will become classical phase and intensity modulations according to these random numbers. Therefore, we can easily conclude that the entanglement-based scheme in Figure \ref{fig:PMencodingClassical} is equivalent to the PM scheme in Box \ref{box:PMQKD}.

Note that the CNOT operation between systems $A_0$ and $B_0$, the quantum Fourier transformation, the photon number measurement, and the $X$-basis measurements on systems $A_1$ and $B_1$ are operations introduced to obtain the quantum phase error rate. These operations were not part of the original protocol and are referred to as virtual operations. In the following sections, we will demonstrate that through these virtual operations, Alice and Bob can obtain the photon number distribution of the pulses they send, and establish a connection between this distribution and the quantum phase error rates.

%In the entanglement picture, Alice and Bob use a qudit ancillary system $A_0, B_0$ to control their discrete random phase, $2\pi {j}/{d}$, in a coherent way, as shown in Figure \ref{fig:PMencoding}. 

\subsection{Phase randomization and pseudo-Fock state}
Firstly, let us focus on one side of the system. It can be calculated that in the encoding circuit depicted in Figure \ref{fig:PMsecurity}, users can directly obtain the photon number in the pulses through coherent operations on ancillary systems, without the need for measurements on the optical systems $A$ or $B$.
%At first, we observe that Alice and Bob can acquire the number of photons via coherent operations.
\begin{observation}
	The number of photons emitted in the optical mode $A(B)$ can be acquired by doing a high-dimensional $Z$-basis measurement after the inverse quantum Fourier transformation on ancillary qudit $A_0(B_0)$ which is used to do the discrete phase randomization.
\end{observation}

Here, we provide an analysis using Alice's side as an example. It's evident from symmetry that on Bob's side, we can arrive at the same conclusion. Concretely, Alice can obtain the photon number information by performing the inverse $d$-dimensional Hadamard gate, or inverse quantum Fourier transformation 
\begin{equation}
	\begin{split}
		F^\dagger &\equiv \sum_{j=0}^{d-1}\ketbra{\tilde{j}}{j}, \\
		\ket{\tilde{j}} &= \frac{1}{\sqrt{d}} \sum_{k=0}^{d-1} e^{2\pi\mathrm{i} jk/d} \ket{k},
	\end{split}
\end{equation} 
on the qudit ancillary system $A_0$ and measuring it in the computational basis,
\begin{equation}
	\begin{split}
		\ket{+_d}_{A_0}\ket{\alpha}_A &\xrightarrow{C\text{-}\theta} \frac{1}{\sqrt{d}} \sum_{j=0}^{d-1} \ket{j}_{A_0}\ket{e^{2\pi\mathrm{i} \frac{j}{d}}\alpha}_A \\
		&\xrightarrow{F^\dagger\otimes I} \frac{1}{\sqrt{d}}\sum_{j=0}^{d-1} \ket{\widetilde{-j}}_{A_0}\ket{e^{2\pi\mathrm{i} \frac{j}{d}}\alpha}_A \\ 
		&\quad \quad= \frac{1}{d} \sum_{j=0}^{d-1} \sum_{k=0}^{d-1} e^{-2\pi\mathrm{i} kj/d}\ket{k}_{A_0} \ket{e^{2\pi\mathrm{i} \frac{j}{d}}\alpha}_A \\
		&\xrightarrow[\text{with outcome }k]{\text{measure }A_0} \sum_{j=0}^{d-1} e^{-2\pi\mathrm{i} kj/d} \ket{e^{2\pi\mathrm{i} \frac{j}{d}}\alpha}_A, \\
	\end{split}
\end{equation}
where the final state is not necessarily normalized after Alice obtains the outcome $k\in[d]$. Here, $\ket{j}$ and $\ket{\tilde{j}}$ represent the eigenstates of the $Z$ and $X$ bases, respectively, in d-dimensional space. We can show that the state in mode $A$ becomes a pseudo-Fock state with photon number $k$,
\begin{equation}\label{pm6:pseudoFock}
	\begin{split}
		\sum_{j=0}^{d-1} e^{-2\pi\mathrm{i} kj/d} \ket{e^{2\pi\mathrm{i} \frac{j}{d}}\alpha} &= e^{-\abs{\alpha}^2} \sum_{j=0}^{d-1} e^{-2\pi\mathrm{i} kj/d} \sum_{n=0}^{\infty} e^{2\pi\mathrm{i} nj/d}\frac{\alpha^n}{\sqrt{n!}}\ket{n} \\
		&= e^{-\abs{\alpha}^2} \sum_{n=0}^{\infty} \left(\sum_{j=0}^{d-1} e^{-2\pi\mathrm{i} kj/d} e^{2\pi\mathrm{i} nj/d}\right) \frac{\alpha^n}{\sqrt{n!}}\ket{n} \\
		&= e^{-\abs{\alpha}^2} \sum_{n=k,k+d,k+2d,\dots} \frac{\alpha^n}{\sqrt{n!}}\ket{n},
	\end{split}
\end{equation}
where in the second equality, the coefficient in the bracket is non-zero only when $n-k$ is a multiplier of $d$. We know that this state is very close to a Fock state $\ket{k}$ when $d\gg \abs{\alpha}^2$ \cite{Cao2015discrete}. Thus, in the entanglement-based picture of the PM scheme, Alice and Bob can measure mode $A_0, B_0$ under the Fourier basis to obtain the photon number in $A$.

In practice, the inversed Fourier transform $F^{\dagger}$ is hard to implement. Fortunately, instead of implementing this operation, we only use this process to analyse the phase error. So the implementation of $F^{\dagger}$ will not be an obstacle to us, as we will discuss in the following section.

\subsection{Total photon number and phase difference}
In the PM scheme, Alice and Bob randomized their phases in optical modes $A$ and $B$, respectively. This is equivalent to measuring $A_0,B_0$ in $Z$ basis. After the detection announcement, they want to compare the random phases for post-selection. From Alice's point of view, her (pseudo) photon number measurement in the Fourier basis is incompatible with the random phase measurement in the $Z$ basis. Fortunately, Alice and Bob only need to know the difference between their random phases. And the phase error is only related to the sum photon number of the two optical modes, as we will discuss later. These two measurements are compatible. This is also the starting point of the symmetry protected security proof \cite{Zeng2019Symmetryprotected} of the PM scheme. In the entanglement-based picture, the users can simultaneously obtain the total photon number and the random phase difference by introducing a high-dimensional CNOT operation between systems $A_0$ and $B_0$, as shown in Figure \ref{fig:PMsecurity}, along with appropriate measurements. 

\begin{observation}
	The total number of photons emitted in two optical modes $A$ and $B$ and the random phase difference can be acquired simultaneously by doing a high-dimensional CNOT operation between systems $A_0$ and $B_0$ followed by a inverse quantum Fourier transformation on ancillary qudit $A_0$ and $Z$-basis measurements on $A_0$ and $B_0$.
\end{observation}

Here, we present the measurement process of total photon number and phase difference, and analyse the post-measurement state.
\begin{equation}\label{eq:phaserandom}
	\begin{split}
		\ket{+_d}_{A_0}\ket{\alpha}_{A}\ket{+_d}_{B_0}\ket{\beta}_{B} &\xrightarrow{C\text{-}\theta} \frac{1}{d} \left(\sum_{j_a=0}^{d-1} \ket{j_a}_{A_0}\ket{e^{2\pi\mathrm{i} \frac{j_a}{d}}\alpha}_A\right) \left(\sum_{j_b=0}^{d-1} \ket{j_b}_{B_0}\ket{e^{2\pi\mathrm{i} \frac{j_b}{d}}\beta}_B \right) \\
		&\xrightarrow[A_0 \text{ to } B_0]{\text{minus}} \frac{1}{d} \sum_{j_a=0}^{d-1}\sum_{j_b=0}^{d-1} \ket{j_a}_{A_0}\ket{e^{2\pi\mathrm{i} \frac{j_a}{d}}\alpha}_A \ket{j_b-j_a}_{B_0}\ket{e^{2\pi\mathrm{i} \frac{j_b}{d}}\beta}_B \\
		&\quad= \frac{1}{d} \sum_{j=0}^{d-1} \sum_{j_a=0}^{d-1} \ket{j_a}_{A_0}\ket{e^{2\pi\mathrm{i} \frac{j_a}{d}}\alpha}_A \ket{j}_{B_0}\ket{e^{2\pi\mathrm{i} \frac{j+j_a}{d}}\beta}_B \\
		&\xrightarrow[\text{on }A_0]{F^\dagger} \frac{1}{d} \sum_{j=0}^{d-1} \sum_{j_a=0}^{d-1} \ket{\widetilde{-j_a}}_{A_0}\ket{e^{2\pi\mathrm{i} \frac{j_a}{d}}\alpha}_A \ket{j}_{B_0}\ket{e^{2\pi\mathrm{i} \frac{j+j_a}{d}}\beta}_B \\
		&\quad= d^{-\frac{3}{2}} \sum_{j=0}^{d-1} \sum_{j_a=0}^{d-1} \sum_{k=0}^{d-1} e^{-2\pi\mathrm{i} kj_a/d}\ket{k}_{A_0} \ket{e^{2\pi\mathrm{i} \frac{j_a}{d}}\alpha}_A \ket{j}_{B_0}\ket{e^{2\pi\mathrm{i} \frac{j+j_a}{d}}\beta}_B \\
		&\xrightarrow[k, j]{\text{measure}} \frac{1}{\sqrt{d}}\sum_{j_a=0}^{d-1} e^{-2\pi\mathrm{i} kj_a/d} \ket{e^{2\pi\mathrm{i} \frac{j_a}{d}}\alpha}_A \ket{e^{2\pi\mathrm{i} \frac{j+j_a}{d}}\beta}_B.
	\end{split}
\end{equation}
where for the ket of mode $B_0$, $\ket{j_b-j_a}$, there is modulo $d$ for the subtraction. We change the variable $j=(j_b - j_a)\mod d$ and then $j_b=(j_a+j)\mod d$. For the phase, modulo $d$ is automatically taken. In the last line, $k$ is the total ``photon number" measured by Alice and $j$ is the random-phase difference measured by Bob.

%\begin{figure}[hbtp!]
%	\begin{quantikz}[row sep={0.9cm,between origins}]
%		\lstick{discrete phase $\ket{+_d}$} & \phase[blue,label position=above]{A_0} & \push{\cdots} & \ctrl{1} & \gate{F^\dagger} & \meterD{Z}  \\
%		\lstick{discrete phase $\ket{+_d}$} & \phase[red,label position=below]{B_0} & \push{\cdots} & \targ{} & \meterD{Z} 
%	\end{quantikz}		
%	\caption{From Alice's and Bob's ancillary systems $A_0$ and $B_0$ for random phases, they can obtain both the phase difference index $j$ and the total photon number $k$. Here, the controlled-minus operation takes $\ket{j_a}\ket{j_b}$ to $\ket{j_a}\ket{(j_b-j_a)\mod d}$.} \label{fig:phaseN}
%\end{figure}

Now, similar to Eq.~\eqref{pm6:pseudoFock}, we can evaluate the (unnormalized) post-measurement state, by denoting $\beta'=e^{2\pi\mathrm{i} \frac{j}{d}}\beta$,
\begin{equation}
	\begin{split}
		\sum_{j_a=0}^{d-1} & e^{-2\pi\mathrm{i} kj_a/d} \ket{e^{2\pi\mathrm{i} \frac{j_a}{d}}\alpha}\ket{e^{2\pi\mathrm{i} \frac{j_a}{d}}\beta'} \\
		&= e^{-(\abs{\alpha}^2+\abs{\beta}^2)/2} \sum_{j_a=0}^{d-1} e^{-2\pi\mathrm{i} kj_a/d} \left(\sum_{n=0}^{\infty} e^{2\pi\mathrm{i} nj_a/d}\frac{\alpha^n}{\sqrt{n!}}\ket{n}\right) \left(\sum_{m=0}^{\infty} e^{2\pi\mathrm{i} mj_a/d}\frac{\beta'^m}{\sqrt{m!}}\ket{m}\right) \\
		&= e^{-(\abs{\alpha}^2+\abs{\beta}^2)/2} \sum_{j_a=0}^{d-1} \sum_{n=0}^{\infty} \sum_{m=0}^{\infty} e^{\frac{2\pi\mathrm{i}}{d}(n+m-k)j_a}  \frac{\alpha^n\beta'^m}{\sqrt{n!m!}}\ket{n} \ket{m} \\
	\end{split}
\end{equation}
Here, we notice that 
\begin{equation}\label{PMs:phasesum0}
	\begin{split}
		\sum_{j_a=0}^{d-1}e^{\frac{-2\pi\mathrm{i}}{d}(n+m-k)j_a} =0,
	\end{split}
\end{equation}
unless $n+m-k$ is a multiplier of $d$. We can change the variable $N=m+n$, denoting the total number of photons in optical modes $A$ and $B$. Then, the (unnormalized) post-measurement state can be written as,
\begin{equation}
	\begin{split}
		\sum_{N} \sum_{m=0}^{N} \frac{\alpha^{N-m}\beta'^m}{\sqrt{(N-m)!m!}}\ket{N-m} \ket{m} \\
	\end{split}
\end{equation}
where the summation of $N$ take values of $N=k,k+d,k+2d,\cdots$. Again, let us take the assumption that $d\gg \abs{\alpha}^2+\abs{\beta}^2$. Then, we can ignore the higher-order terms in the pseudo Fock states, $N=k+d,k+2d,\cdots$. That is, we can directly set $N\approx k$,
\begin{equation}
	\begin{split}
		\sum_{m=0}^{k} \frac{\alpha^{k-m}\beta'^m}{\sqrt{(k-m)!m!}}\ket{k-m} \ket{m}, \\
	\end{split}
\end{equation}
which is a Fock state in two modes.

In practice, Alice will not perform the inversed Fourier transform to measure the total photon number. Thus, we can remove $F^{\dagger}$ operation in practice. Then we can move the $Z$ basis measurement to the front of the controlled minus. The highly entangled controlled minus operation becomes classical. Furthermore, the $Z$ basis measurement can be moved previous to the controlled phase operation, making the controlled phase operation classical. This is equivalent to that Alice and Bob first randomly pick a phase $\phi_a, \phi_b$, and modulate the phase of the coherent state. Then, they compare their phase $\phi_a, \phi_b$ to post-select the key. This process reduces the entanglement picture to the original protocol.

We point out that this reduction will not affect our security proof. In principle, Alice can perform the total photon number measurement. As we will prove in the next section, the state after total photon number measurement contains the secure part and the insecure part. The security is only related to the parity of the total photon number and is irrelevant to the measurement result $k$. Thus, whether or not Alice measures $k$, the secure part exists. The entanglement picture derives an upper bound on the phase error rate of the original protocol. In addition, Alice and Bob do not carry out the virtual operations in actual experiments. They only perform $Z$-basis measurements on the systems $A_1$ and $B_1$ to obtain the raw key bits and further estimate the quantum bit error rate. Therefore, in our entanglement picture, users can always upper bound the quantum phase error rate while obtaining the quantum bit error rate.

\subsection{Phase-matching scheme with key encoding}
Based on the case discussed in the former section, we now include the key-bit encoding in the scheme. As shown in Fig.~\ref{fig:PMsecurity}, Alice applies another qubit ancillary system $A_1$. She prepares $\ket{+}$ on $A_1$ and employs a controlled-phase gate between $A_1$ and $A$ to encode the key information. Bob applies similar operations as well. Then, the state becomes
\begin{equation}
	\begin{split}
		&\ket{+_d}_{A_0}\ket{+}_{A_1}\ket{\alpha}_{A}\ket{+_d}_{B_0}\ket{+}_{B_1}\ket{\beta}_{B} \\
		&\xrightarrow{C\text{-}\theta_j} \frac{1}{d} \left(\sum_{j_a=0}^{d-1} \ket{j_a}_{A_0}\ket{+}_{A_1}\ket{e^{2\pi\mathrm{i} \frac{j_a}{d}}\alpha}_A\right) \left(\sum_{j_b=0}^{d-1} \ket{j_b}_{B_0}\ket{+}_{B_1}\ket{e^{2\pi\mathrm{i} \frac{j_b}{d}}\beta}_B \right) \\
		&\xrightarrow[C\text{-}\theta_b]{C\text{-}\theta_a}\frac{1}{2d} \left[\sum_{j_a=0}^{d-1} \ket{j_a}_{A_0}\left(\ket{0}_{A_1}\ket{e^{2\pi\mathrm{i} \frac{j_a}{d}}\alpha}_A+\ket{1}_{A_1}\ket{-e^{2\pi\mathrm{i} \frac{j_a}{d}}\alpha}_A\right)\right] \left[\sum_{j_b=0}^{d-1} \ket{j_b}_{B_0}\left(\ket{0}_{B_1}\ket{e^{2\pi\mathrm{i} \frac{j_b}{d}}\beta}_B+\ket{1}_{B_1}\ket{-e^{2\pi\mathrm{i} \frac{j_b}{d}}\beta}_B\right) \right]\\
		&\xrightarrow[A_0 \text{ to } B_0]{C\text{-}X}\frac{1}{2d} \left[\sum_{j_a=0}^{d-1} \ket{j_a}_{A_0}\left(\ket{0}_{A_1}\ket{e^{2\pi\mathrm{i} \frac{j_a}{d}}\alpha}_A+\ket{1}_{A_1}\ket{-e^{2\pi\mathrm{i} \frac{j_a}{d}}\alpha}_A\right)\right] \left[\sum_{j_b=0}^{d-1} \ket{j_b-j_a}_{B_0}\left(\ket{0}_{B_1}\ket{e^{2\pi\mathrm{i} \frac{j_b}{d}}\beta}_B+\ket{1}_{B_1}\ket{-e^{2\pi\mathrm{i} \frac{j_b}{d}}\beta}_B\right) \right]\\
		&\quad=\frac{1}{2d} \left[\sum_{j=0}^{d-1} \sum_{j_a=0}^{d-1}\ket{j_a}_{A_0}\left(\ket{0}_{A_1}\ket{e^{2\pi\mathrm{i} \frac{j_a}{d}}\alpha}_A+\ket{1}_{A_1}\ket{-e^{2\pi\mathrm{i} \frac{j_a}{d}}\alpha}_A\right)\ket{j}_{B_0}\left(\ket{0}_{B_1}\ket{e^{2\pi\mathrm{i} \frac{j + j_a}{d}}\beta}_B+\ket{1}_{B_1}\ket{-e^{2\pi\mathrm{i} \frac{j + j_a}{d}}\beta}_B\right) \right]\\
		&\xrightarrow[\text{on }A_0]{F^\dagger}\frac{1}{2d} \left[\sum_{j=0}^{d-1} \sum_{j_a=0}^{d-1}\ket{\widetilde{-j_a}}_{A_0}\left(\ket{0}_{A_1}\ket{e^{2\pi\mathrm{i} \frac{j_a}{d}}\alpha}_A+\ket{1}_{A_1}\ket{-e^{2\pi\mathrm{i} \frac{j_a}{d}}\alpha}_A\right)\ket{j}_{B_0}\left(\ket{0}_{B_1}\ket{e^{2\pi\mathrm{i} \frac{j + j_a}{d}}\beta}_B+\ket{1}_{B_1}\ket{-e^{2\pi\mathrm{i} \frac{j + j_a}{d}}\beta}_B\right) \right]\\
		&\quad= \frac{1}{2d^{\frac{3}{2}}} \sum_{j=0}^{d-1} \sum_{j_a=0}^{d-1} \sum_{k=0}^{d-1} e^{-2\pi\mathrm{i} kj_a/d}\ket{k}_{A_0} \left(\ket{0}_{A_1}\ket{e^{2\pi\mathrm{i} \frac{j_a}{d}}\alpha}_A+\ket{1}_{A_1}\ket{-e^{2\pi\mathrm{i} \frac{j_a}{d}}\alpha}_A\right)\ket{j}_{B_0}\left(\ket{0}_{B_1}\ket{e^{2\pi\mathrm{i} \frac{j + j_a}{d}}\beta}_B+\ket{1}_{B_1}\ket{-e^{2\pi\mathrm{i} \frac{j + j_a}{d}}\beta}_B\right) \\
		&\xrightarrow[k,j]{\text{measure}} \sum_{j_a=0}^{d-1} e^{-2\pi\mathrm{i} kj_a/d}\left(\ket{0}_{A_1}\ket{e^{2\pi\mathrm{i} \frac{j_a}{d}}\alpha}_1+\ket{1}_{A_1}\ket{-e^{2\pi\mathrm{i} \frac{j_a}{d}}\alpha}_A\right) \left(\ket{0}_{B_1}\ket{e^{2\pi\mathrm{i} \frac{j_a}{d}}\beta'}_B+\ket{1}_{B_1}\ket{-e^{2\pi\mathrm{i} \frac{j_a}{d}}\beta'}_B\right),
	\end{split}
\end{equation}
where for the ket of mode $B_0$, $\ket{j_b-j_a}$, there is modulo $d$ for the subtraction. We change the variable $j=(j_b - j_a)\mod d$ and $\beta'=e^{2\pi\mathrm{i} \frac{j}{d}}\beta$.
%where $j_b\ominus j_a$ takes subtraction with modulo $d$. We change the variable $j=j_b\ominus j_a$ and $\beta'=e^{2\pi\mathrm{i} \frac{j}{d}}\beta$. 
This is the same as Eq.~\eqref{eq:phaserandom}. Similarly, we can evaluate the (unnormalized) post-measure state,
\begin{equation}\label{PMs:kFock}
	\begin{split}
		&\sum_{j_a=0}^{d-1} e^{-2\pi\mathrm{i} kj_a/d}\left(\ket{0}_{A_1}\ket{e^{2\pi\mathrm{i} \frac{j_a}{d}}\alpha}_A+\ket{1}_{A_1}\ket{-e^{2\pi\mathrm{i} \frac{j_a}{d}}\alpha}_A\right) \left(\ket{0}_{B_1}\ket{e^{2\pi\mathrm{i} \frac{j_a}{d}}\beta'}_B+\ket{1}_{B_1}\ket{-e^{2\pi\mathrm{i} \frac{j_a}{d}}\beta'}_B\right) \\
		&\quad=e^{-(\abs{\alpha}^2+\abs{\beta}^2)/2} \sum_{j_a=0}^{d-1} e^{-2\pi\mathrm{i} kj_a/d} \\
		&\quad[\ket{00}_{A_1B_1}\left(\sum_{n=0}^{\infty} e^{2\pi\mathrm{i} nj_a/d}\frac{\alpha^n}{\sqrt{n!}}\ket{n}_A\right) \left(\sum_{m=0}^{\infty} e^{2\pi\mathrm{i} mj_a/d}\frac{\beta'^m}{\sqrt{m!}}\ket{m}_B\right)\\
		&\quad+\ket{01}_{A_1B_1}\left(\sum_{n=0}^{\infty} e^{2\pi\mathrm{i} nj_a/d}\frac{\alpha^n}{\sqrt{n!}}\ket{n}_A\right) \left(\sum_{m=0}^{\infty} e^{2\pi\mathrm{i} mj_a/d}\frac{(-\beta')^m}{\sqrt{m!}}\ket{m}_B\right)\\
		&\quad+\ket{10}_{A_1B_1}\left(\sum_{n=0}^{\infty} e^{2\pi\mathrm{i} nj_a/d}\frac{(-\alpha)^n}{\sqrt{n!}}\ket{n}_A\right) \left(\sum_{m=0}^{\infty} e^{2\pi\mathrm{i} mj_a/d}\frac{\beta'^m}{\sqrt{m!}}\ket{m}_B\right)\\
		&\quad\ket{11}_{A_1B_1}\left(\sum_{n=0}^{\infty} e^{2\pi\mathrm{i} nj_a/d}\frac{(-\alpha)^n}{\sqrt{n!}}\ket{n}_A\right) \left(\sum_{m=0}^{\infty} e^{2\pi\mathrm{i} mj_a/d}\frac{(-\beta')^m}{\sqrt{m!}}\ket{m}_B\right)]\\
		&\quad = e^{-(\abs{\alpha}^2+\abs{\beta}^2)/2} \sum_{j_a=0}^{d-1}\sum_{n=0}^{\infty}\sum_{m=0}^{\infty}e^{\frac{2\pi\mathrm{i}}{d}(n+m-k)j_a} 
		\left[\ket{0}+(-1)^{n}\ket{1}\right]_{A_1}\left[\ket{0}+(-1)^{m}\ket{1}\right]_{B_1} \frac{\alpha^n\beta'^m}{\sqrt{n!m!}}\ket{nm}_{AB}.
	\end{split}
\end{equation}
Here, we change the order of some systems for simplicity and use the subscript to denote the system. Note that according to Eq.~\eqref{PMs:phasesum0}, we have $N=n+m=k, k+d, k+2d, \cdots$. The phase error is obtained by $X \otimes X$ measurement on the key qubits $A_1, B_1$. Then, we can determine whether there exists a phase error by the total photon number. 
%According to Eq.~\eqref{PMs:kFock}, we have the following results for different cases based on the parity of $N$.
\begin{enumerate}
	\item 
	$N$ is odd, the $X$-basis measurement results on qubits $A_1$ and $B_1$ are different, $\ket{-+}_{A_1B_1}$ or $\ket{+-}_{A_1B_1}$;
	\item 
	$N$ is even, the $X$-basis measurement results on qubits $A_1$ and $B_1$ are the same $\ket{++}_{A_1B_1}$ or $\ket{--}_{A_1B_1}$.
	%\item 
	%$N$ is odd, $n$ is odd, and $m$ is even. The $X$-basis measurement results on qubits $A_1$ and $B_1$ are different, $\ket{-+}_{A_1B_1}$;
	%
	%\item 
	%$N$ is odd, $n$ is even, and $m$ is odd. The $X$-basis measurement results on qubits $A_1$ and $B_1$ are different, $\ket{+-}_{A_1B_1}$;
	%
	%\item 
	%$N$ is even, $n$ is even, and $m$ is even. The $X$-basis measurement results on qubits $A_1$ and $B_1$ are the same $\ket{++}_{A_1B_1}$;
	%
	%\item 
	%$N$ is even, $n$ is odd, and $m$ is odd. The $X$-basis measurement results on qubits $A_1$ and $B_1$ are the same, $\ket{--}_{A_1B_1}$.
\end{enumerate}
In conclusion, the phase error rate is $1$ if the total photon number in $A$ and $B$ is odd, and it is $0$ if the total photon number is even. Then the upper bound of the phase error rate is
\begin{equation}
\begin{split}
	e_p&\leq 1\cdot q_\text{even} + 0\cdot q_\text{odd}\\
	&=q_\text{even} \\
	&=1-\sum_k q_{2k+1}.
\end{split}
\end{equation}
This is consistent with Eq.~\eqref{eq:epupper} in Sec.~\ref{sc:symmetrypro}. As previously discussed, our result can be applied to the original protocol. Thus, by analyzing the fraction of odd and even state via decoy state method\cite{Ma2005practical}, the upper bound of phase error can be derived. Utilizing the entanglement-based source-replacement picture, we establish the relationship between the total photon number and the phase error rate in the PM scheme. This provides a more concrete and physically intuitive explanation for the conclusions drawn in the symmetry-protected privacy method.

\section{Beam-splitting attacks}\label{sc:attack}
From the security analyses presented in Sec.~\ref{sc:preliminary} and \ref{sc:SourceRep}, it is apparent that the conclusions drawn from the source-replacement and symmetry-protected privacy security analyses are entirely consistent. These two approaches appear to be equivalent to some extent, but their starting points are not identical. One is based on the entire source and the quantum states it emits, while the other is based on encoding operations. Therefore, in practical applications, it is essential to choose the appropriate analysis method based on specific requirements.

In this section, we introduce a new attack mainly based on the beam-splitting attack and the unambiguous state discrimination approach and attempt to analyze the key rate of the PM scheme under this attack. Similar attack strategies were initially introduced in Ref.~\cite{ferenczi2013security} to target MDI quantum key distribution protocols. Here, we draw inspiration from these attack ideas and apply our approach to analyze the phase error rate under this attack. In this scenario, we find that the security analysis approach based on source replacement can be directly applied to analyze the attack and straightforwardly establish a lower bound on the phase error rate. This shows the advantages of the source-replacement security analysis approach in analyzing attacks.

\subsection{Attack strategy}

\begin{figure}[!htbp]
	\centering \resizebox{8cm}{!}{\includegraphics{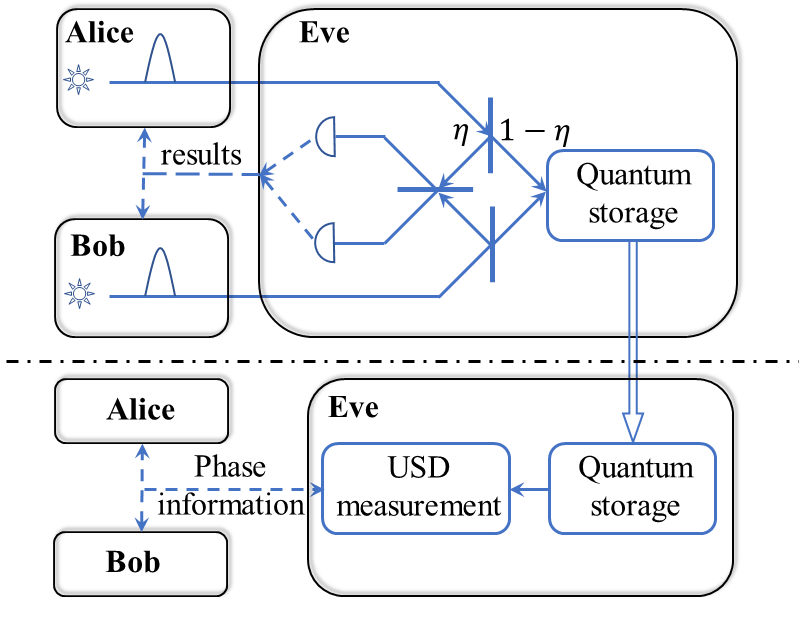}}
	\caption{Illustration of beam-splitting attack. Solid arrows represent the transmission of quantum states, while dashed arrows represent the exchange of classical information. The upper part of the figure corresponds to step (1) in Box \ref{box:BSattack}. Eve splits the light pulse emitted by Alice and Bob into two parts with a ratio of $1-\eta:\eta$. The former part is stored in a quantum memory, while the latter part undergoes interference measurement, and the measurement result is publicly announced. The lower part of the figure corresponds to steps (2) and (3). After the measurement is completed and Alice and Bob announce the phase information $\phi^i_a$ and $\phi^i_b$, Eve retrieves the two corresponding states from the quantum memory, performs unambiguous state discrimination measurement based on the phase information, and applies post-processing to the results. Between these two stages, the only transmitted quantum states are the states held by Eve in the quantum memory.} \label{fig:attack}
\end{figure}

Firstly, we give the detailed description of the proposed attack in Box \ref{box:BSattack}, and we also illustrate this attack in Figure \ref{fig:attack}. Here we suppose the channel transmittance from Alice and Bob to the measurement site are both $\eta$, and the intensities of pulses are the same, $\mu_a=\mu_b=\mu$. We only consider the case of pure states for simplicity.

\begin{mybox}[label={box:BSattack}]{Beam-splitting attack strategy}%\label{box:BB84}
	\begin{enumerate}[(1)]
		
		\item 
		Beam splitting and measurement: In the $i$-th round, as soon as Alice and Bob send out the quantum state, $\ket{\alpha_i},\ket{\beta_i}$, Eve immediately intercepts the light pulses and passes the intercepted pulses through a beamsplitter with a transmission coefficient of $\eta$. From the reflection port, Eve captures and stores the weak coherent states that emerge, denoted as $\ket{\varphi_i}=\ket{\sqrt{(1-\eta)\mu}e^{i\theta_a}}\ket{\sqrt{(1-\eta)\mu}e^{i\theta_b}}$, in a quantum memory. Here, $\theta_a=\pi\kappa^i_a+\phi^i_a$ and $\theta_b=\pi\kappa^i_b+\phi^i_b$ represent the encoded phases, which are unknown to Eve. For the pulses that emerge from the transmission port, Eve proceeds to perform a single-photon interference measurement and publicly announces the measurement results, same as the procedure carried out by Charlie in steps (2) and (3) described in Box \ref{box:PMQKD}.
		
		\item
		Announcement: For those rounds with successful detection, Alice and Bob announce the random phases, $\phi^i_a$ and $\phi_b^i$.
		
		\item %\label{StepAnnounce}
		Unambiguous state discrimination: 
		After Alice and Bob announce the random phases, Eve extracts the states $\ket{\varphi_i}$ stored in the quantum memory and attempts to distinguish the encoded state using the unambiguous state discrimination method. Without loss of generality, we assume they sift $\phi^i_a = \phi_b^i$. Otherwise, Eve could perform an additional $\pi$ phase flip on one of the states in her possession. Then, Eve only need to distinguish two states $\ket{\varphi_i^0}$ and $\ket{\varphi_i^1}$ to get the key bit,
		where $\ket{\varphi_i^0} = \ket{\sqrt{(1-\eta)\mu}}\ket{\sqrt{(1-\eta)\mu}}$ and $\ket{\varphi_i^1} =\ket{-\sqrt{(1-\eta)\mu}}\ket{-\sqrt{(1-\eta)\mu}}$. 
		
		%After Alice and Bob repeat their steps for $N$ times, they will post-select rounds with successful detection and $\abs{\phi^i_a-\phi^i_b}=0$ or $\pi$, and map the key. For these rounds, Eve extracts the states $\ket{\varphi_i}$ stored in the quantum memory and attempts to distinguish which encoding state each of these states corresponds to using the unambiguous state discrimination method. She then determines the eavesdropped key bits according to the measurement results. 
		
		%\item
		%Data postprocessing: Finally, Eve will obtain a eavesdropped key string that is the same length as the raw key string held by Alice and Bob. Subsequently, Eve performs the corresponding operations based on the information reconciliation and privacy amplification procedures conducted by Alice and Bob on their raw key strings. 
		
	\end{enumerate}
\end{mybox}

The core idea of this beam-splitting attack lies in that the state $\ket{\varphi}$ obtained by Eve through beam splitting is very close to the states emitted by Alice and Bob. Moreover, as the channel transmission rate decreases, these two states become even closer. With the help of quantum memory, Eve can utilize the stored states to attempt to obtain the key bits chosen by Alice and Bob, given that she knows the random phases selected by them. In this scenario, Eve can maximize the utilization of all the information she can obtain without interfering with the protocol execution, thus maximizing her ability to steal the key. Whether Eve can obtain the key information or not depends on her ability to distinguish whether the state she holds is $\ket{\varphi^0}$ or $\ket{\varphi^1}$ through unambiguous state discrimination measurements. If she successfully distinguishes these two states, Eve will perfectly learn the key bit value. The optimal success probability for her to distinguish these two states is given by the fidelity between these two states \cite{JAEGER199583}, 
\begin{equation}
	p_{usd} = 1-\abs{\braket{\phi_0}{\phi_1}} = 1-e^{-4(1-\eta)\mu}.
\end{equation}

\subsection{Phase error rate estimation and simulation}
Given the probability for Eve to perfectly learn the key bit value, we can further estimate a lower bound on the phase error rate that Alice and Bob will encounter in the protocol. To illustrate this point more clearly, we still use the source-replacement entanglement-based picture, as shown in Figure \ref{fig:PMsecurity}. In this picture, the states of systems $A_1$ and $B_1$ are the key states held by Alice and Bob, respectively. They will perform $Z$-basis measurements on their own states to obtain their raw keys. According to the definition of the phase error rate \cite{lo1999Unconditional}, if they perform $X$-basis measurements on these states, they will obtain the phase error rate. 

When Eve successfully obtains Alice and Bob's encoded keys through unambiguous state discrimination measurements, in the entanglement-based scenario, we can consider that Eve deterministically acquired knowledge of the results of the $Z$-basis measurements performed on systems $A_1$ and $B_1$. In this sense, the states on these two systems have already collapsed to either $\ket{00}$ or $\ket{11}$ from Eve's point of view.
%When Eve successfully obtains Alice and Bob's encoded keys through unambiguous state discrimination measurements, in the entanglement-based scenario, we can consider that 
%Eve hacks the systems $A_1$ and $B_1$ and carries out $Z$-basis measurements on these two systems. 
Then Alice and Bob do subsequent operations to get the raw key bits or the phase error rate. If Alice and Bob perform $Z$-basis measurements, the raw key bits they get will be the same as what Eve obtained. If Alice and Bob perform $X$-basis measurements to estimate the phase error rate, since systems $A_1$ and $B_1$ have already collapsed to either $\ket{00}$ or $\ket{11}$, the results of the $X$-basis measurements will be completely random, resulting in a phase error rate of $\frac12$. This scenario here is similar to the intercept-resend attack on the BB84 protocol. 
Therefore, the contribution of the case where Eve successfully distinguishes the encoded states to the phase error rate is $\frac12p_{usd}$. As for the case where Eve fails to distinguish the states, the phase error rate is lower bounded by $0$ since any additional operation will only increase the phase error rate.
Hence, we can conclude that the final phase error rate satisfies 
\begin{equation}
	e_p \ge \frac12 p_{usd} = \frac12 - \frac12 e^{-4(1-\eta)\mu} \equiv e_p^L.
\end{equation}
Note that although we have only considered the case of pure states here, in the case of mixed states, we can always assume that Eve holds a purification of the mixed state. Therefore, she can only make $p_{usd}$ better, so it is a valid lower bound.
Then, by substituting the upper and lower bounds of the phase error rates obtained into the key rate formula, we can derive the upper bound on the key rate obtained from the beam-splitting attack and the lower bound on the key rate obtained from symmetry-protected security proof, respectively.

\begin{figure*}[htbp]
	\centering
	\resizebox{12cm}{!}{\includegraphics{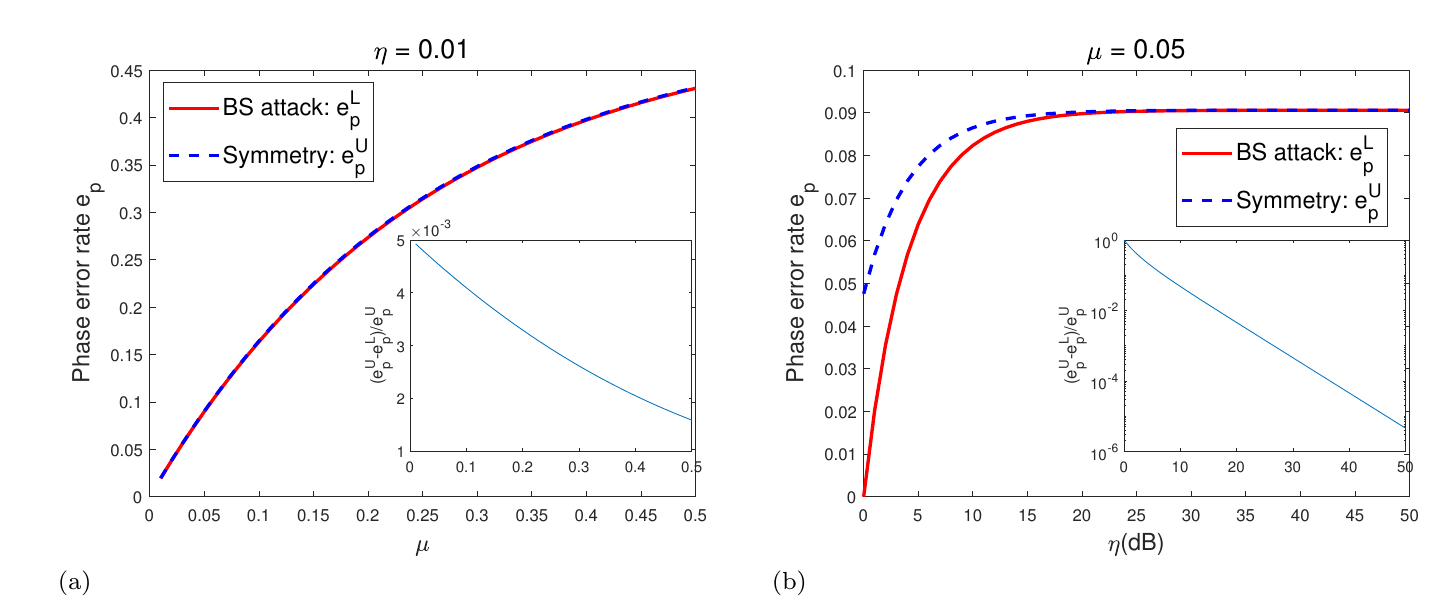}}
	\caption{The difference between the phase error rate upper bounded by the symmetry-protected security proof and the beam-splitting attack under different intensities and channel transmittance. In addition to the specific values of the phase error rate given by the two methods, we also included a small plot in each figure showing the ratio of the difference between them. (a) In this figure, we set the channel transmittance to be 0.01, corresponding to a communication distance of 200 km from Alice to Bob using standard optical fibers with an attenuation of 0.2 dB/km. For the optical intensity, we selected a range from 0 to 0.5, which is commonly used in QKD protocols. (b) In this figure, we set the optical intensity to be 0.05, which is a typical sending intensity of phase-matching QKD \cite{fang2020implementation}. For the channel transmittance, we selected a range from 0 dB to 50 dB, corresponding to communication distances ranging from 0 km to 500 km with standard optical fibers. } \label{fig:epgap}
\end{figure*}

To visually illustrate the difference in phase error rates obtained from these two methods, we conducted simulations of the phase error rates for various optical intensities and channel transmittance. We also calculated the ratio of the difference, $\dfrac{e_p^u-e_p^L}{e_p^u}$ between the two rates. The results are presented in Figure \ref{fig:epgap}.

The figures show that the upper and lower bounds of the phase error rate given by the two methods are very close. The difference becomes smaller as the intensity increases and the channel transmission decreases. Intuitively, these results are also reasonable. As the intensity increases or the channel transmission rate decreases, Eve can obtain quantum states with higher intensities through beam splitting, which makes the states Eve holds closer to the ones sent by Alice and Bob. And it will finally lead to a higher success probability of unambiguous state discrimination. If the communication distance between Alice and Bob is zero, Eve cannot obtain any quantum states through beam splitting. 
We need to emphasize that a typical sending intensity of phase-matching QKD is in the order of $10^{-2}$ \cite{fang2020implementation}, implying that our simulation results encompass the practical scenarios of phase-matching QKD in practical implementations. The results indicate that the lower bounds on key rates are highly tight, especially for long communication distances, leaving little room for further enhancement. Similarly, the results can also imply that the attack we proposed, along with the corresponding calculation and analysis method for the phase error rate, provides a good lower bound under the beam-splitting attack. The analysis process also demonstrates that the entanglement-based security analysis approach can provide straightforward conclusions with simplicity when analyzing such attacks. This highlights the advantages of the entanglement-based security analysis method when dealing with these types of attacks.

\section{Conclusion and outlook}\label{sc:conclusion}
In this work, we reexamine the security of the PM scheme and sought a more tangible and intuitive understanding of its security. Our approach introduced a source-replacement model, offering a perspective on the PM scheme's security based on entanglement. By employing this source-replacement method, we revealed that it yielded the same conclusions as the symmetry-protected privacy method. Specifically, both approaches establish that quantum states with an odd total photon number result in phase errors, while states with an even total photon number remain phase-error-free.

Furthermore, we extended our analysis to include potential threats in the form of a beam-splitting attack scheme. By considering the entanglement properties associated with this attack, we quantified its impact on the protocol's performance, providing an upper bound on the phase error rate. Our analysis proves the tightness of the key rate and establishes a direct link between attacks and quantum phase error rates. This approach is also applicable to other QKD protocols, including continuous-variable ones that have attracted considerable attention from the quantum communication community. 

Note that our current source-replacement security analysis method can mainly be applied to phase-based encoding protocols. It cannot directly provide conclusions consistent with the symmetry-protected privacy method for intensity-based encoding protocols like the mode-paring scheme \cite{zeng2022mode} and sending-or-not-sending twin-field scheme \cite{wang2018twin}. How to extend this source-replacement picture to such protocols and prove their security is an intriguing topic, and we hope that our findings will inspire further research and development in the field of QKD.
Furthermore, we also hope that our proposed beam-splitting attack and its analysis can inspire the exploration of new attack strategies, the investigation of the security of different QKD protocols, and the development of more robust and rigorous security analysis methods. By enhancing protocols and strengthening security measures, we hope it will be helpful to advance the practicality and security of QKD systems, paving the way for their widespread adoption in real-world applications.
%Furthermore, our analysis provides a new perspective for examining the relationship between QKD security proofs and attacks. We hope that our findings will inspire further research and development in the field of QKD, like exploring new attack strategies, investigating the security of different QKD protocols, and developing more robust and tight security analysis methods. By enhancing protocols and strengthening security measures, we hope it will be helpful to advance the practicality and security of QKD systems, paving the way for their widespread adoption in real-world applications.

\section*{Acknowledgments}
This work was supported by the National Natural Science Foundation of China Grant No.~12174216 and the Innovation Program for Quantum Science and Technology Grant N0.~2021ZD0300804.
%\section*{Competing Interests}
%The authors declare no competing interests.
%%%%%%%%%%%%%%%%%%%%%%%%%%%%%%%%%%%%%%%%
% choose a .bib file
\bibliography{./tex/bibPM.bib}
%%%%%%%%%%%%%%%%%%%%%%%%%%%%%%%%%%%%%%%%
\end{document}